\documentclass[twocolumn,twoside,slac_two]{revtex4}
\usepackage{graphicx}
\usepackage{fancyhdr}
\begin{document}

\title{$\gamma$-ray Flaring Emission in Blazar OJ287 Located in the Jet $\gtrsim14$\,pc from the
Black Hole}

%

\author{I. Agudo$^{1,2}$, S. G.~Jorstad$^{2,3}$, A. P.~Marscher$^{2}$, V. M.~Larionov$^{3,4}$, J. L.~G\'omez$^{1}$, A. L\"{a}hteenm\"{a}ki$^{5}$, M. Gurwell$^{6}$, P. S. Smith$^{7}$, H. Wiesemeyer$^{8,9}$, C. Thum$^{10}$ and J. Heidt$^{11}$}
\affiliation{{$^{1}$Instituto de Astrof\'{i}sica de Andaluc\'{i}a, CSIC, Granada, Spain}\\
{$^{2}$Institute for Astrophysical Research, Boston University,Boston, USA}\\
{$^{3}$Astronomical Institute, St. Petersburg State University, St. Petersburg, Russia}\\
{$^{4}$Isaac Newton Institute of Chile, St. Petersburg Branch, St. Petersburg, Russia}\\
{$^{5}$Aalto University Mets\"{a}hovi Radio Observatory, Kylm\"{a}l\"{a}, Finland}\\
{$^{6}$Harvard--Smithsonian Center for Astrophysics, Cambridge, USA}\\
{$^{7}$Steward Observatory, University of Arizona, Tucson, USA}\\
{$^{8}$Instituto de Radio Astronom\'{i}a Milim\'{e}trica, Granada, Spain}\\
{$^{9}$Max-Plack-Institut f\"ur Radioastronomie, Bonn, Germany}\\
{$^{10}$Institut de Radio Astronomie Millim\'{e}trique, St. Martin d'H\`{e}res, France}\\ 
{$^{11}$ZAH, Landessternwarte Heidelberg, K\"{o}nigstuhl, Heidelberg, Germany}
}

\begin{abstract}
We combine the Fermi-LAT light curve of the BL Lacertae type blazar OJ287 with time-dependent multi-waveband flux and linear polarization observations and submilliarcsecond-scale polarimetric images at $\lambda= 7$\,mm to locate the
$\gamma$-ray emission in prominent flares in the jet of the source $\gtrsim14$\,pc from the central engine. 
We demonstrate a highly significant correlation between the strongest  $\gamma$-ray and millimeter-wave flares through Monte Carlo simulations.
The two reported $\gamma$-ray peaks occurred near the beginning of two major millimeter-wave outbursts, each of which is associated with a linear polarization maximum at millimeter wavelengths. 
Our very long baseline array observations indicate that the two millimeter-wave flares originated in the second of two
features in the jet that are separated by $>14$\,pc. 
The simultaneity of the peak of the higher-amplitude $\gamma$-ray flare and the maximum in polarization of the second jet feature implies that the $\gamma$-ray and millimeter-wave flares are cospatial and occur $\gtrsim14$\,pc from the central engine. 
We also associate two optical flares, accompanied by sharp polarization peaks, with the two $\gamma$-ray
events. 
The multi-waveband behavior is most easily explained if the $\gamma$-rays arise from synchrotron self Compton scattering of optical photons from the flares. 
We propose that flares are triggered by interaction of moving plasma blobs with a standing shock. 
\end{abstract}

\maketitle

\thispagestyle{fancy}


\section{Introduction}
 The Large Area Telescope (LAT) onboard the {\it Fermi} Gamma-ray Space Telescope has sufficient sensitivity to study the location of the $\gamma$-ray emission site in blazars through well-sampled light curves that allow detailed studies of the timing of $\gamma$-ray flares relative to those at other spectral ranges \citep[e.g.,][]{Abdo:2010p11811,Jorstad:2010p11830,Marscher:2010p11374}.
The technique developed by \citet{Marscher:2010p11374} and \citet{Jorstad:2010p11830} uses ultra-high angular-resolution monitoring with VLBI to resolve the innermost jet regions and monitor changes in jet structure. 
Monthly observations, provide time sequences of total and polarized intensity images of the parsec-scale jet that can be related to variations of the flux and polarization at higher frequencies. 
In \citet{Agudo:2011p14707}, we employed this technique to investigate the location of the flaring $\gamma$-ray emission in the BL~Lacertae (BL~Lac) object {OJ287} ($z=0.306$).
In this paper we reproduce the results in \citet{Agudo:2011p14707}. 
We adopt a cosmology with $H_0$=71 km s$^{-1}$ Mpc$^{-1}$, $\Omega_M=0.27$, and $\Omega_\Lambda=0.73$, so that 1~mas corresponds to a projected distance of 4.48~pc, and a proper motion of 1\,mas/yr corresponds to a superluminal speed of $19\,c$.

\section{Observations}
\begin{figure}
   \centering
   \includegraphics[clip,width=5.2cm]{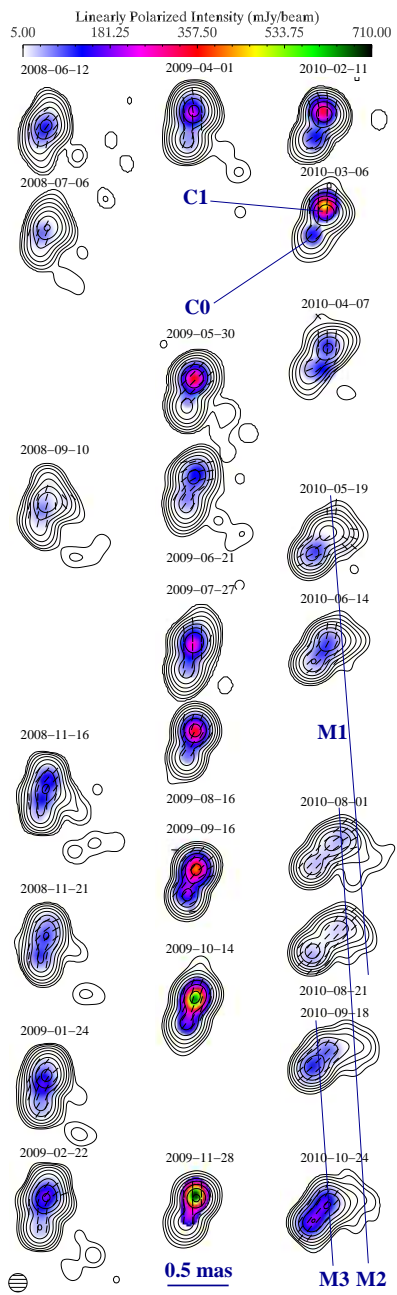}
   \caption{Sequence of 7\,mm VLBA images of OJ287 in 2008--2010.
   Images are convolved with a $\rm{FWHM}=0.15$\,mas circular Gaussian beam.
   Contour levels represent 0.2, 0.4, 0.8, 1.6, 3.2, 6.4, 12.8, 25.6, 51.2, 90.0\% of the peak total intensity of 6.32\,Jy/beam. 
   The color scale indicates linearly polarized intensity, whereas superimposed line segments represent the orientation of the polarization electric-vector position angle.}
   \label{maps}
\end{figure}

Our polarimetric observations of {OJ287} include (1) 7\,mm VLB) images (Fig.~\ref{maps}), mostly from the Boston University blazar monitoring program, (2) 3\,mm monitoring with the IRAM 30\,m Telescope, and (3) optical ($R$ and $V$ band) photo-polarimetric observations from several observatories (Figs.~\ref{tflux} and \ref{pol}).  
The optical facilities include Calar Alto (2.2\,m telescope, observations under the MAPCAT program), Steward (2.3 and 1.54\,m telescopes), Lowell (1.83\,m Perkins Telescope), St. Petersburg State University (0.4\,m telescope), and Crimean Astrophysical  (0.7\,m telescope) observatories. 
To these we add $R$-band polarimetric (and $V$-band photometric) data from \citet{Villforth:2010p11557}.
The total flux light curves analyzed here (see Fig.~\ref{tflux}) are from the \emph{Fermi}-LAT $\gamma$-ray (0.1--200\,GeV) and \emph{Swift} X-ray (0.3--10\,keV) and optical ($V$-band) data available from the archives of these missions, by the Yale University SMARTS program, by the Submillimeter Array (SMA) at 1.3\,mm and 850\,$\mu$m, by the IRAM 30\,m Telescope at 1.3\,mm, and by the Mets\"{a}hovi Radio Observatory 14\,m Telescope at 8\,mm.

Our data reduction follows: (1) VLBA: \citet{Jorstad:2005p264,Agudo:2007p132}; (2) optical polarimetric data: \citet{Jorstad:2010p11830,Larionov:2008p338}; (3) IRAM data: \citet{Agudo:2006p203,Agudo:2010p12104,Agudo:2011p15946}; (4) SMA: \citet{Gurwell:2007p12057}; (5) Mets\"{a}hovi: \citet{1998A&AS_132_305T}; (6) \emph{Swift}: \citet{Agudo:2011p14707}; and (7) \emph{Fermi} LAT: \citet{Agudo:2011p14707}. 

\section{Flares in the C1 Jet Region at 1\,mm and 7\,mm}
We model the brightness distribution in the 7\,mm VLBA images of the source (Fig.~\ref{maps}) with a small number of circular Gaussian components.
Our model fits include a bright quasi-stationary feature (C1) $\sim0.2$\,mas from the innermost jet region (C0, that we identify with the 7\,mm core \citep{Agudo:2011p14707}).

\begin{figure}
   \centering
   \includegraphics[clip,width=7.cm]{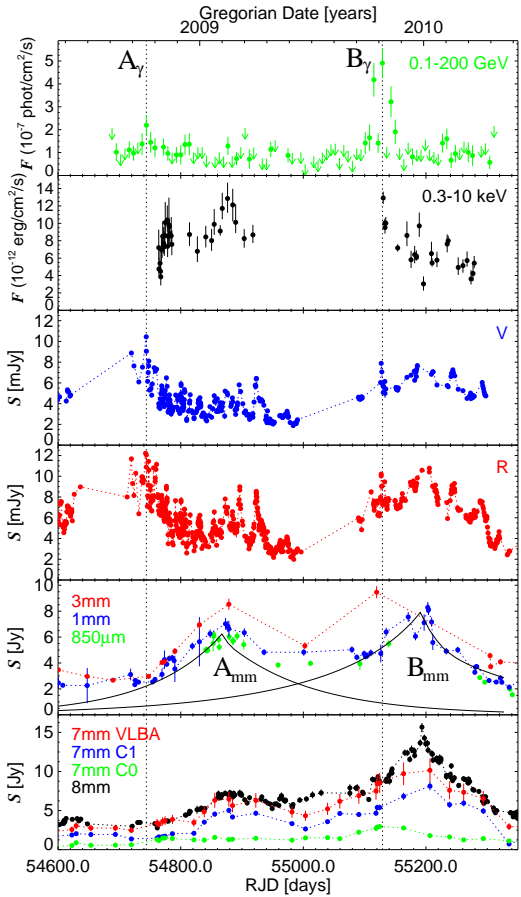}
   \caption{Light curves of OJ287 from mm-wave to $\gamma$-ray frequencies. The vertical lines denote the times of peak $\gamma$-ray flux of ${\rm{A}}_{\gamma}$ and ${\rm{B}}_{\gamma}$. RJD = Julian Date $- 2400000.0$.}
   \label{tflux}
\end{figure}

C0 and C1 typically governed the mm-wave evolution, of {OJ287}.
In particular, the two most prominent 1\,mm flares ever reported in {OJ287} (${\rm{A}}_{\rm{mm}}$ and ${\rm{B}}_{\rm{mm}}$ as labeled in Fig.~\ref{tflux}) took place in C1, as indicated by the correspondence of events in the 7\,mm light curve of this component with those at other millimeter wavelengths.

From the angle of the jet axis to the line of sight in {OJ287} \citep[$1^\circ\kern-.35em .9$--$4^\circ\kern-.35em .1$; see][]{Jorstad:2005p264,Pushkarev:2009p9412}, and the mean projected separation of C1 from C0 at the time of start of ${\rm{A}}_{\rm{mm}}$ and ${\rm{B}}_{\rm{mm}}$ ($0.23\pm0.01$\,mas), we estimate that C1 is located $>14$\,pc downstream of C0, the innermost jet region detected in our images.
The actual distance between the central engine in {OJ287} and C1 must be even greater if C0 lies downstream of the acceleration and collimation zone of the jet \citep[ACZ;][]{2007AJ.134.799J,2008Natur_452_966M}.

\section{$\gamma$-Ray Flares}
\begin{figure}
   \centering
   \includegraphics[clip,width=6.cm]{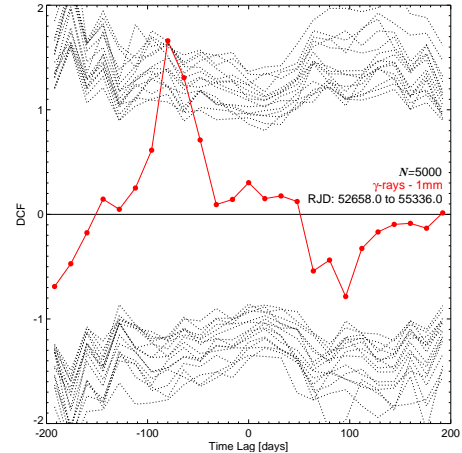}
   \caption{Discrete cross-correlation function between the $\gamma$-ray and 1\,mm light curves of OJ287 (red points). 
   The dotted curves at positive (negative) DCF values denote 99.7\% confidence limits for correlation rather than stochastic variability (see \citet{Agudo:2011p14707} for details).}
   \label{dcf}
\end{figure}

The two most pronounced $\gamma$-ray flares in OJ287 (Fig.~\ref{tflux}) take place during the initial rising phases of ${\rm{A}}_{\rm{mm}}$ and ${\rm{B}}_{\rm{mm}}$. 
Our discrete correlation function (DCF) study between the $\gamma$-ray and 1\,mm long-term light curve -- see \citet{Agudo:2011p14707} for details -- possesses a significant ($>99.7$\,\% confidence) peak at a time lag $\sim -80$ days ($\gamma$-ray leading, see Fig.~\ref{dcf}).
{\it This evidences the correlation between ${\rm{B}}_{\gamma}$ and ${\rm{B}}_{\rm{mm}}$, the most luminous $\gamma$-ray and 1\,mm flares in our data}.

The optical light curves (especially $V$-band; Fig.~\ref{tflux}) show two sharp flux increases at essentially zero time lag from ${\rm{A}}_{\gamma}$ and ${\rm{B}}_{\gamma}$.

\section{Variability of Linear Polarization}
\begin{figure}
   \centering
   \includegraphics[clip,width=7.cm]{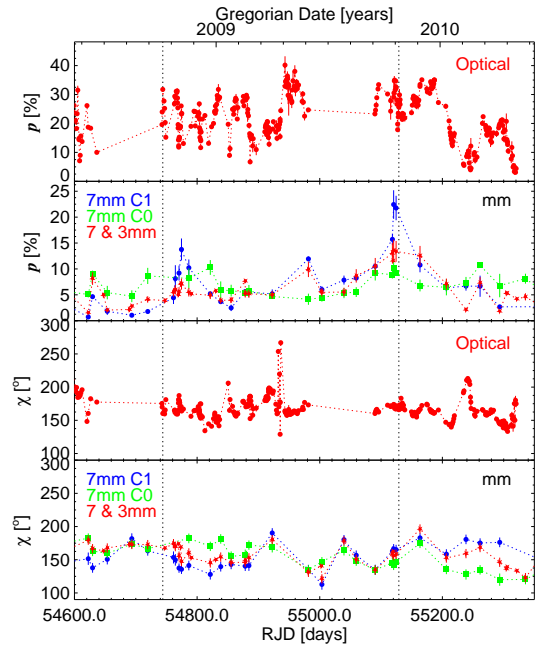}
   \caption{Optical and mm-wave linear polarization of OJ287 as a function of time. The optical data includes R-band and 5000--7000\,{\AA} observations.  The similarity of the integrated linear polarization at 7\,mm and 3\,mm also allows us to combine them. Vertical lines are as in Fig.~\ref{tflux}.}
   \label{pol}
\end{figure}

C0 and C1 dominate the evolution of the linear polarization $p$ and electric vector position angle $\chi$ at 7\,mm in {OJ287} (Fig.~\ref{pol}).  
However, whereas $p_{\rm{C0}}$ never exceeds $10$\%, C1 exhibits the two largest peaks in $p$ ever observed in {OJ287} at 7\,mm, $p_{\rm{C1}}\approx14$\% on 2008 November 4, and $p_{\rm{C1}}\approx22$\% on 2009 October 16.
The first maximum in $p_{\rm{C1}}$ follows the peak of ${\rm{A}}_{\gamma}$ by one month, while the second more pronounced polarization event is already in progress when the $\gamma$-ray flux of flare ${\rm{B}}_{\gamma}$ rises to a level of $\sim 4\times 10^{-7}$~phot~cm$^{-2}~s^{-1}$. 
{\emph{This coincidence of the strongest $\gamma$-ray outburst and exceptionally strong polarization in C1 identifies this feature $>14$\,pc from the central engine as the site of the variable $\gamma$-ray emission.}}

The optical polarization peaks at essentially the same time as $p_{\rm{C1}}$ at 7\,mm during flare ${\rm{B}}_{\gamma}$. 
During both ${\rm{A}}_{\gamma}$ and ${\rm{B}}_{\gamma}$, $p_{\rm{opt}} \approx 35\%$, which requires a well-ordered magnetic field.
However, comparable optical polarization levels also occur at other times.
The shorter time scale and larger amplitude of variability of optical polarization, as compared with those at millimeter wavelengths, is consistent with frequency dependence in the turbulence model of \citet{Marscher:2010p12402}.

The optical and mm-wave linear polarization position angle is stable at $\chi\approx160^{\circ}$--$170^{\circ}$ ($-20^\circ$ to $-10^\circ$)---similar to the structural position angle of the inner jet---both near ${\rm{A}}_{\gamma}$ and ${\rm{B}}_{\gamma}$ and throughout most of the monitoring period. 
The corresponding direction of the magnetic field is transverse to the direction between features C0 and C1.

\section{Discussion and Conclusions}

The two 0.1--200\,GeV flares in {OJ287} allow us to assess the correspondence between $\gamma$-ray and lower-frequency variations. We find that two kinds of events at millimeter wavelengths are related to these $\gamma$-ray outbursts at high significance:
(1) the early, rising phases of the two most luminous 1\,mm flares ever detected in this blazar, (${\rm{A}}_{\rm{mm}}$ and ${\rm{B}}_{\rm{mm}}$); and
(2) two sharp increases to unprecedented levels of linear polarization ($\sim14$\% and $\sim22$\%) in bright jet feature C1 $>14$\,pc from the central engine.
These events also coincide with:
(3) two sharp optical flares;
(4) two peaks in optical polarization of $\sim35$\%; and
(5) the similarity of optical and mm-wave polarization position angle both during and between the flares at $\chi\approx160^{\circ}$-$170^{\circ}$.

The exceptionally high polarization of C1 during $\gamma$-ray flare ${\rm{B}}_{\gamma}$ provides extremely strong evidence that the event occurred in C1. 
The $\gamma$-ray IC emission arises, in principle, from either the synchrotron self-Compton (SSC) process or IC scattering of infrared radiation from a hot, dusty torus of size $\sim 10$~pc (IC/dust). 
IC scattering from the broad line region is not possible given the long distance of the $\gamma$-ray dissipation region that we measure.
An SSC model is possible given the low ratio of $\gamma$-ray to synchrotron luminosity ($\approx 2$ in {OJ287}, see \citet{Agudo:2011p14707}).
However, infrared emission from the dusty torus has not been detected thus far in BL~Lacs such as {OJ287}, as far as the authors know.
We thus favor the SSC mechanism (see \citet{Agudo:2011p14707} for further arguments).

The observed behavior agrees with a scenario in which the optical and $\gamma$-ray flares are produced in C1 by particle acceleration in a moving blob when it crosses a standing shock well beyond the ACZ (see \citet{Agudo:2011p14707} for a sketch). 
\citet{2007ApJ_659L_107D} have associated the innermost mm-wave jet emission feature with a recollimation shock at the end of the ACZ. 
We identify C0 as such a feature, and C1 as a second re-collimation shock, as seen in hydrodynamical simulations \citep{1997ApJ_482L_33G,Agudo:2001p460} and often in VLBA images \citep{Jorstad:2005p264}.

The blob is a turbulent plasma disturbance (typical weakness of the observed polarization outside the high polarization peaks) that propagates down the jet with significantly higher relativistic electron density and magnetic field than in the ambient flow. 
\citet{Cawthorne:2006p409} shows that, in this case, the side of the conical shock nearest to the line of sight can be highly polarized with $\chi$ parallel to the jet axis. 
The remainder of the conical shock, farther from the line of sight, has much lower polarization. 
Because of light-travel delays, we first see the blob penetrate the near side, and therefore observe a major increase in polarization. 
As the outburst develops, more of the emission comes from the low-polarization far side, which decreases $p$ while the mm-wave flux density continues to increase. 
This is the pattern observed during both flares.

The mm-wave and optical flares start at essentially the same time as the magnetic field and electron energies near the leading edge of the blob become amplified as they pass the standing shock front. 
The mm-wave flux outburst continues as the relatively low-energy electrons fill the shocked region. 
The optical and $\gamma$-ray emission, produced by higher-energy electrons that cannot travel far before suffering radiative energy losses, is confined closer to the shock front where particles are accelerated.

\bigskip 

\begin{acknowledgements}
This research was funded by NASA grants NNX08AJ64G, NNX08AU02G, NNX08AV61G, and NNX08AV65G, NSF grant AST-0907893, and NRAO award GSSP07-0009 (Boston University); RFBR grant 09-02-00092 (St.~Petersburg State University); MICIIN grants AYA2007-67627-C03-03 and AYA2010-14844, and CEIC (Andaluc\'{i}a) grant P09-FQM-4784 (IAA-CSIC); the Academy of Finland
(Mets\"{a}hovi); and NASA grants NNX08AW56S and NNX09AU10G (Steward Observatory).
The VLBA is an instrument of the NRAO, a facility of the NSF operated under cooperative agreement by AUI. 
The PRISM camera at Lowell Observatory was developed by Janes et al., with funding from the NSF, Boston University, and Lowell Observatory. The Calar Alto Observatory is jointly operated by MPIA and IAA-CSIC. 
The IRAM 30\,m Telescope is supported by INSU/CNRS (France), MPG (Germany), and IGN (Spain).
The Submillimeter Array is a joint project between the SAO and the Academia Sinica. 
\end{acknowledgements}

\bigskip 

%
%
%
%

\end{document}